# Using Serious Games to Train Evacuation Behaviour

João Ribeiro[1], João Emílio Almeida[1,Ä], Rosaldo J. F. Rossetti[1,Ä], António Coelho[1,Œ], António Leça Coelho[2]

[1]Department of Informatics Engineering
[Ä]LIACC – Laboratory of Artificial Intelligence and Computer Science
[Œ]INESC TEC – INESC Technology and Science
Faculty of Engineering, University of Porto
Rua Roberto Frias, S/N, 4200-465, Porto, Portugal
{joao.pedro.ribeiro, joao.emilio.almeida, rossetti, acoelho}@fe.up.pt

[2]LNEC – National Laboratory of Civil Engineering
Av. Brasil, 101, 1700-066, Lisboa, Portugal
alcoelho@lnec.pt

*Abstract*: **Emergency evacuation plans and evacuation drills are mandatory in public buildings in many countries. Their importance is considerable when it comes to guarantee safety and protection during a crisis. However, sometimes discrepancies arise between the goals of the plan and its outcomes, because people find it hard to take them very seriously, or due to the financial and time resources required. Serious games are a possible solution to tackle this problem. They have been successfully applied in different areas such as health care and education, since they can simulate an environment/task quite accurately, making them a practical alternative to real-life simulations. This paper presents a serious game developed using Unity3D to recreate a virtual fire evacuation training tool. The prototype application was deployed which allowed the validation by user testing. A sample of 30 individuals tested the evacuating scenario, having to leave the building during a fire in the shortest time possible. Results have shown that users effectively end up learning some evacuation procedures from the activity, even if only to look for emergency signs indicating the best evacuation paths. It was also evidenced that users with higher video game experience had a significantly better performance.**

## I. Introduction

Everyone who has gone through an emergency drill exercise has surely noticed some of its downsides. It is hard to take them seriously, to fully participate, to keep focused on the exercise and to learn all the necessary information in order to evacuate correctly. The disruption of the regular functioning of an enterprise or institution easily upsets their occupants, and so they tend to distrust these exercises. For the same reasons, and additionally due to the employees need to move away from their workstations, it is not usual for managers to accept and face fire drills very easily, who generally take this kind of exercises as a loss of time.

A virtual simulation system is a possible alternative to emergency drills, and can indeed produce some useful information. To accomplish a good virtual simulation environment, it is required to provide a realistic scenario and immersion feeling, to make users acquire important, real-world knowledge. This software approach can be useful in order to study possible outcomes to unusual situations and the behaviour of whole populations, but do not supply much interaction; therefore, they can rapidly become uninteresting for users, instead of helping them as intended. Through the use of video games, these difficulties can be addressed and minimised. Video games by definition involve and require user interaction. They are often quite immersive due to their multimedia nature and by creating feedback through video, audio and sometimes even haptic feedback. Video games usually present some characteristics that make them an interesting resource: users stay interested in its execution, they are challenged to master some skills and their performance is evaluated. These applications have recently been used with positive results in several and distinct fields of study, what makes them an option to consider.

The attitude towards them has also changed dramatically: although previously seen as a "waste of time" and "highly addictive", their use in education and training has increased in the past years. It should be noted that video games have been a major driving force in the development of computer technology. They now provide realistic scenarios, forces and interactions, and also user friendly interfaces. Besides, they run on regular home computers, greatly reducing the necessary investment in order to make use of them.

In this paper we explore the concept of serious games to build a virtual fire drill simulator as an attempt to address some of the issues we have identified within real-world fire drills. It is our intention to improve the way people participate in such experiments enhancing their experience in many different ways. We have adapted and custom the environment of a game engine, in this case Unity3D, to support simulation features that enable users to be tracked and assessed while playing.

The remaining of this paper is organised as follows. In Section II we briefly present some related concepts of subjects that concern this project, such as serious games, pedestrian simulation, and crowd behavioural modelling. We then discuss on applying serious games to evacuation training, in Section III, where we present and formalise our problem and propose the approach implemented in this paper. A preliminary experiment using our prototype is presented in Section IV, where some results are also discussed in Section V. We finally draw some conclusions in Section VI and suggest some further steps in this research.

## II. BACKGROUND

Before we can go further in the explanation of our approach to create a virtual environment to aid and enhance traditional fire drills, it is important to introduce some valuable concepts and some basics of subjects that underlie the implementation of this project.

### A. Serious Games

According to Hays [1], a game is an "artificially constructed, competitive activity with a specific goal, a set of rules and constraints that is located in a specific context." Despite their origins being based purely with entertainment in mind, video games have been recently used to achieve other goals and objectives. Serious Games refer to video games whose application is focused on supporting activities such as education, training, health, advertising, or social change.

Video games present some characteristics that make them helpful as a resource. As sentiments like enjoyment and fulfilment are felt, users are more likely to stay motivated while doing their tasks. Freitas states in [2] a few benefits from combining them with other training activities:

- the learners' motivation is higher;
- completion rates are higher;
- possibility of accepting new learners;
- possibility of creating collaborative activities;
- learn through doing and acquiring experience.

Other aspects that draw video game players' attention are fantasy elements, challenging situations and the ability to keep them curious about the outcomes of their possible actions [3].

According to experts from the European Centre for Children's Products [4], Serious Games can be classified in five categories: Edutainment, Advergaming, Edumarket Games, Political Games and Training and Simulation Games. Each of these categories can find their practical application in a diverse range or areas, proving to be an invaluable asset and instrument for training and influencing behavioural patterns.

### B. Pedestrian Simulation

Pedestrian Simulation is a field of computing that has gained a considerable dimension more recently. Its applications lean mostly towards planning urban areas, studying and testing the safety of buildings or open spaces and evaluating emergency systems and evacuation routes. Other uses include crowd animation for films or video games, as well.

This field of study involves different technology areas, as Artificial Intelligence and Crowd Simulation, but also Decision Making and Behaviour Analysis. One major issue the scientific community faces though is how to accurately model and implement human behaviour and feature simulated entities with rather real attitudes and reactions. This has been addressed through many different approaches, mainly based on psychological experiments. Even so, it is also hard to keep subjects motivated and committed to this kind of experiments. In this area too, serious games seem to have a great potential as a research tool, fostering the understanding of human behaviour and the decision-making process carried out by individuals in a variety of circumstances.

### C. Crowd Representation Models

Although many approaches exist to virtually simulate the behaviour of crowds with varying levels of realism and computational efficiency, from the analysis of different sources [5, 6, 7] three models seem to be the most used:

- **Cellular Automata Models** - treats individuals as separate objects within a space that is discretely represented through a matrix of cells. Besides allowing only discrete movement and positioning of the elements in the space, with high-density crowds the base individual areas are exposed. Even so, it can still be a valid option as it demands lower computational resources and still provides reasonably realistic crowd behaviours macroscopically.
- **Forces-based Models** - use mathematical formulas to calculate the position variations of the individual elements through the application of forces. Its most explored subtypes consider Magnetic Forces and Social Forces. The former model assigns polarities to individuals and objects and calculates attractions, directions and movement based on these values. As for the Social Forces approach, it is based on the first one and draws on its concept of polarity to manage individuals' objectives and wishes.
- **Artificial Intelligence-based Models** - the decisions are taken by the individual elements that compose the crowd. One of the most common subtypes of this model is the Agent-based Model, whose unit is the Agent. Agents are elements capable of individually deciding their actions according to their own perceptions and interactions. If used in a high enough number, these models can be a good approximation to a crowd representation. Most importantly, agent-based models are a good metaphor for representing social interactions and behaviour, which is of paramount importance in crowd modelling and simulation.

### D. Crowd Behaviour

Groups of individuals present different behaviour patterns under normal and emergency events [8, 9]. One important factor that has great influence over people's movement and actions is the layout of a building or surrounding areas.

Helbing [8] considers that the following attributes should be accounted for and definitely included in the representation of each virtual individual to simulate their behaviour:

- **State** - including its position, health and mobility;

- **Speed** - possible moving speeds;
- **Vision** - field of vision, influenced by smoke;
- **Reaction Time** - elapsed time before changing from normal to evacuation behaviour;
- **Collaboration** - influences the probability of helping other people;
- **Insistence** - probability of maintaining the same objective over time;
- **Knowledge** - familiarity with the building and the necessary actions to take to evacuate.

Concerning the simulation of large groups under stress because of an emergency situation, Cordeiro [10] provides some insight into additional possible effects such situations might have on the group. It has been observed, for instance, that "herding" (or "flocking") happens due to people following other individuals instinctively (a reaction thoroughly explored by Reynolds in [11]), believing that they will be able to evacuate faster and more effectively. However, in conditions of low visibility or little knowledge of the surroundings this can provoke flocks of wandering people, contributing to the panic and confusion of the whole group, which is also a social reaction rather to be avoided if possible.

III. USING SERIOUS GAMES IN EVACUATION TRAINING

Whereas pedestrian and crowd simulations are an important instrument to help us better understanding social and group behaviour in a vast number of different circumstances, they are not good enough to persuade people to behave correctly. On the other hand, traditional fire drills as they are implemented in reality albeit constituting a very useful training resource have also some drawbacks to be sorted out. In our approach, we combine pedestrian and crowd simulation with the concept of serious games to build upon an aid tool that can be used to influence subjects' behaviour towards a more effective instinctive reaction when responding to stressing situations such as the need to evacuate their current environment.

*A. Problem definition*

Performing real simulation exercises is an important activity nowadays, as it helps buildings' users train evacuation routes, learn useful behaviours and prepare themselves in case there is an emergency situation. Having the occupants of a building well trained in the necessary procedures can help prevent casualties in case of an uncommon event (which is even more relevant for large, crowded buildings and ones that receive many visitors). In the end of a fire drill, it is expected to have identified some problems with the building's safety routes and attempt to correct them. These simulations create the conditions to implement recent methods for assessing buildings' safety levels (such as Performance Based Design - PBD), replacing more traditional methods that present the adequate solution for a given situation. In some countries they are mandatory and even required at certain intervals.

Despite this, due to being uncommon these exercises can be extremely disruptive to the usual activities and operation of an institution, more so when they are unannounced. A briefing must be provided so everyone knows what is going to happen, how they should proceed and what is expected from them all. They are required to abandon their workstations for some moments, halt their work and leave the building before returning to their places and resuming their activities. Alarm mechanisms are frequently activated for the duration of the exercise, emergency authorities are often required to be present and evacuation times are measured. For management executives, who also have to perform these activities, all the trouble and derangement might be seen as a complete loss of time and a factor disrupting production.

In these simulations, it is almost impossible for someone to grasp all the necessary information in order to evacuate from a building. One might remember later which path was taken previously, but there are also other numerous details, ranging from building plans to all possible evacuation routes and gather points. Thus, it is difficult to retain all this knowledge and have it present in times of real necessity.

It should also be considered that in order to maintain high levels of realism in these activities a high investment must be made to recreate threats, hazards and uncommon events. Objects that represent fire and smoke machines can be used, for instance. Nonetheless, unpredictable hazards cannot be fully recreated, due to their dangerous nature. This is a typical example to which dealing directly with the real environment can ultimately compromise its normal functioning.

Some simulation software, known as egress or evacuation simulators, has been used in recent years to solve part of the difficulties aforementioned. Indeed they can be useful to substitute some of the real activities, by providing a less expensive alternative, which also benefits from being more easily started, planned and executed. They might be shown to several occupants at once, minimising the time needed for the exercise, and can be of help in identifying best routes for leaving the building, finding bottlenecks and obstacles, and passing general information. Nevertheless, their aim is usually more to provide theoretical feedback on the current situation of a building regarding evacuation. They can be of use for a building in project, in order to correctly plan exits, paths, and illumination, or for existing buildings so as to evaluate how its population could behave in case of an emergency and unexpected hazardous events. Even the highly realistic ones do not provide an immersive experience to the users, as they cannot be put them in charge of controlling a virtual person of that simulation.

*B. The proposed approach*

Video games can be a choice to consider in order to address the aforementioned issues. Their potential to make tedious tasks become enjoyable is considerable and they have been recently applied to different ends other than solely providing entertainment to their users.

As McGonigal presents it [12], it takes a clear goal that we can achieve and direct feedback for work to feel positive and even pleasant, and games excel at providing both. In fact, the author even states that by only bestowing our best efforts into overcoming a game challenge, we feel more rewarded.

According to Davidson [13], a key quality of video games is that they focus not only on learning and knowing some information but actually on using it appropriately. One other aspect that is relevant is that video games are useful by providing an experience not far from real life with lower cost and without compromising safety [14]. Hazards can also be controlled more easily to provide randomness to the simulations. This project's main objective is to study the viability of using a game engine to create an application that is simultaneously a pedestrian simulator, a game and a fire evacuation training tool.

*C. Implementation*

A Game Engine is a set of tools that provide necessary features to the development of video games. The required features in our specific situation as listed below:

- 3D renderer;
- Audio playback;
- Physics engine;
- Graphical User Interfaces (GUI);
- Collision detection;
- Support for artificial intelligence.

After comparing various game engines, Unity3D was selected. In fact, it presents among other features a powerful interface that allows visual object placement and property changing during runtime (especially useful to rapidly create new scenarios from existing models and assets and quick tweaking of script variables). The framework is also customisable, giving the developer the ability to create code in JavaScript, C# or Boo. Finally, it also provides a simple project deployment environment for multiple platforms, with no need for additional configuration, including the web (which makes it possible to run any game on a web browser).

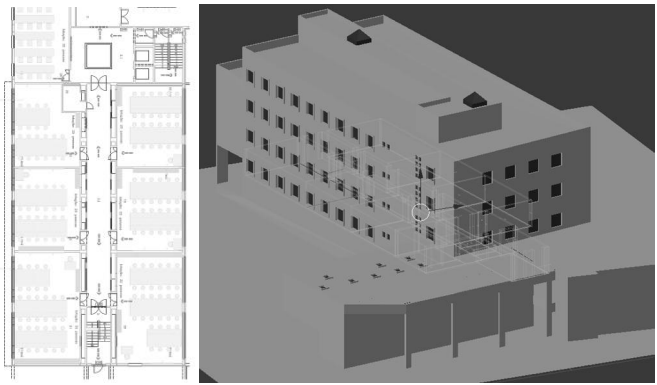

Figure 1.  DEI plan and 3D model

Currently, our framework contains a single simulation scenario. It takes place in FEUP's Informatics Engineering Department (DEI). The model was created using Blender based on the actual blueprints of the building so as to recreate it as realistically as possible in terms of topology, dimensions, scale and proportions Figure 1. The 3D environment can be seen in Figure 2, which gives a good idea of how realistic such an environment can represent the actual place under study.

*D. Execution of a evacuation scenario*

The player starts in a predefined room and, upon starting the evacuation event, a fire appears in a random room, forcing the alarm to ring. At this moment, the timer starts and the player must then traverse the building in order to go outside as quickly as possible, choosing from one of the two possible exits. Several emergency signs are in place so as to help the player to identify the nearest exit to his/her current spot.

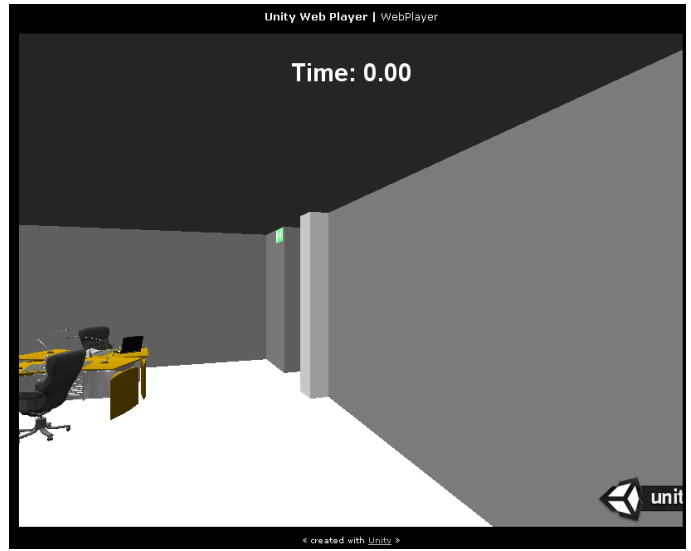

Figure 2.  3D rendered view of the building

This attempts to recreate the experience of the user being physically there and exploring their surroundings. To steer his/her digital avatar in a 3D virtual world the user has the controls that are commonly standard for the FPS game genre (First Person Shooter games characterised by placing players in a 3D virtual world which is seen through the eyes of an avatar).

IV. EVALUATION OF PRELIMINARY RESULTS

In order to test our approach and evaluate its potentials, a simple experiment was executed and some results analysed. Our findings have been quite encouraging and promising, demonstrating the viability of our methodology.

*A. Population Sample*

A total of 30 individuals participated in this experiment and were invited to test the developed prototype. Testers can be individually classified according to different dimensions, namely whether they are regular video game players and whether they had previous knowledge of the building. According to these characteristics, four groups were considered. The different groups were labelled A, B, C and D as follows:

- A - previous building knowledge and frequent game players;
- B - previous building knowledge and little experience with games;

- C - no previous building knowledge and frequent game players;
- D - no previous building knowledge and little experience with games;

The distribution of our sample into each of the groups previously described is presented in Table 1.

TABLE I. USER GROUPS

| Group | Frequent game player | Previous building knowledge | Population |
|---|---|---|---|
| A | Yes | Yes | 8 |
| B | No | Yes | 6 |
| C | Yes | No | 5 |
| D | No | No | 11 |

*B. Test Setup*

Before the evacuation test started, all players had a few moments to try the controls and get used to them (this was done without the possibility of exploring the building or even leaving the starting room). Players with little to no familiarity with the building additionally had a small demonstration of the way from the main entrance to the room where the test would start, to have in mind at least one way out.

All users had to perform the same task in order to generate comparable results. Starting in a room inside DEI, the player's objective was then to reach the outside of the building in the shortest time possible. The alarm would sound and computer-controlled agents would also start to move towards the nearest exit, crowding the corridors and causing the natural disruptions we can usually find when facing this kind of situations. With time, fire would spread all over the place and become an even more serious obstacle. While navigating inside the building, players could refer to the emergency signs pointing to the nearest emergency exits.

The only measured value was the time taken to evacuate the building, as mistakes made while finding the way out would be automatically reflected in a longer elapsed time.

## V. RESULTS AND DISCUSSION

While carrying out our experiments, some performance measures were logged for each subject. Figure 3 depicts the times for each player.

Very curiously, the average time to leave the building was higher than expected. After all, in real life it only takes around 22 seconds to do the same way. This might have been originated by the difficulties verified by some users to control the avatar, mostly ones who demonstrated little to no acquaintance with video games on a regular basis. Some of them required several seconds to get used to controlling mouse and keyboard at the same time; in some cases they even gave up trying to do it simultaneously and opted instead by alternating between pointing at the desired direction and then moving on that way.

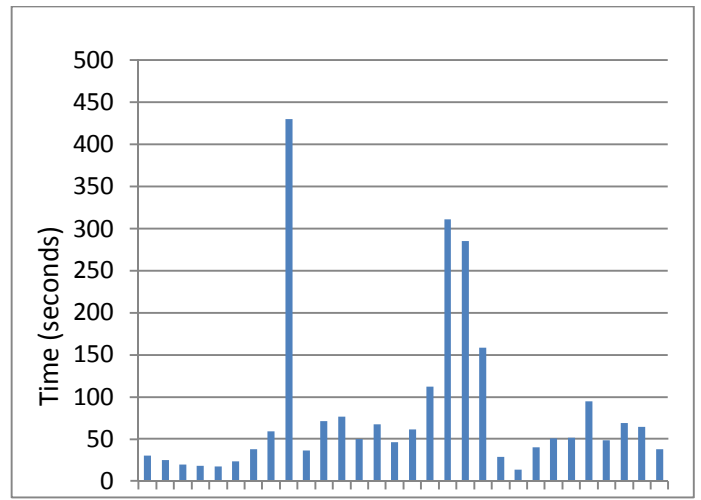

Figure 3. User times for each player

Average evacuation times computed by category are shown in Table II, below.

TABLE II. AVERAGE TIMES BY CATEGORY

| Group | Average time (seconds) |
|---|---|
| A | 23.9 |
| B | 43.9 |
| C | 58.0 |
| D | 145.1 |

It is not surprising that the group with better performance was that of regular video game players who were also familiar with the building (group A). It was also expected to see the opposite group – group D, people that are neither familiar with DEI's building nor regular gamers – have the lowest score. It is the two middle scores that deserve more attention.

Not knowing the building or not being used to the game controls: both groups present similar results. Whilst the frequent game players where faster in controlling the avatar, they missed the shortest path to outside and choose the usual way due to their poor building knowledge (group C); the group with little game experience but good building familiarity (group B) although using the emergency exit took more time due to little experience with games. Frey et al. [15] have even shown that by making users get used to a game, the difficulties with controls are quickly overcome and the subjects differences on this point mitigated. To the benefit of the experience, players were allowed to play only once. When repeating the game, all would select the emergency exit and evacuation times decreased substantially, so results would be biased. It was precisely the novelty of facing such a new situation that was the aim of this experience.

## VI. Conclusions and Future Work

In this work we explore the concept of serious games as an important asset to aid and improve traditional fire drills. It does neither completely replace nor avoid the need for in-site drills to train people in emergency situations, such as with the prospect of fire in an office building, for instance. Nonetheless, game environments can be very attractive in many different ways, and have proven to be an invaluable tool for training. Additionally, we build our approach on the potential of such a concept to ease and improve the understanding of human behaviour in such situations, as subjects are monitored during their playing the game and some performance measures are logged to be further analysed later on.

We have implemented our prototype with Unity3D, a popular game engine, which provided us with a customisable framework and allowed us to insert features of a serious game platform into the virtual environment. We invited some subjects to use the game and collected some preliminary results that demonstrate the viability of the approach. We have then conceived a methodology which is both instrumental as an aid to train people and an invaluable instrument to help practitioners and scientists to better understand group behaviour and the social phenomenon in a vast range of circumstances.

The very next steps in this research include the improvement of the prototype featuring it with tools for rapidly setting up simulation environments from CAD blueprints of buildings. We also intend to include other performance measures to study individual and social behaviour in circumstances other the hazardous scenarios. Ultimately, this tool is also expected to be used as an imperative decision support tool, providing necessary and additional insights into evacuation plans, building layouts, and other design criteria to enhance places where people usually gather and interact rather socially, such as shopping malls, stadiums, airports, and so on.

## ACKNOWLEDGMENT

This project has been partially supported by FCT (*Fundação para a Ciência e a Tecnologia*), the Portuguese Agency for R&D, under grant SFRH/BD/72946/2010.